# Positive Quantum Brownian Evolution

Allan Tameshtit[1] and J. E. Sipe[2]

[1]Department of Physics, Harvard University, Cambridge, Massachusetts 02138
[2]Department of Physics, and Ontario Laser and Lightwave Research Centre, University of Toronto,
Toronto, Ontario, Canada M5S-1A7

(Submitted 10 June 1996)

Using the independent oscillator model with an arbitrary system potential, we derive a quantum Brownian equation assuming a correlated total initial state. Although not of Lindblad form, the equation preserves positivity of the density operator on a restricted set of initial states.

Brownian motion is both ubiquitous in nature and of fundamental significance in the study of systems that interact with their surroundings. At a classical level where the environment is often idealized as a set of harmonic oscillators, the phenomenon is well understood. Yet a description at the quantum mechanical level, even within this independent oscillator model of the environment, is still problematic. In particular, consider the standard quantum Brownian equation in the literature describing the evolution of the reduced density operator of the system, $\rho(t)$ [1,2]:

$$\frac{d\rho(t)}{dt} = \frac{1}{\hbar i}[H', \rho(t)] - \frac{CkT}{2\hbar}\{q, \rho(t), q\} - \frac{Ci}{8m}\{p, \rho(t), q\} + \frac{Ci}{8m}\{q, \rho(t), p\}, \quad (1)$$

where $T$ is the temperature of the surroundings, $C$ is a constant measuring the strength of the environmental coupling, $H'$ is the system Hamiltonian plus Hermitian terms proportional to $C$, and $\{A, \rho, B\} \equiv BA^{\dagger}\rho + \rho BA^{\dagger} - 2A^{\dagger}\rho B$. Equation (1) is known to be ill behaved [3–8]; it does not in general preserve the positivity of $\rho(t)$; i.e., the condition that $\langle\psi|\rho(t)|\psi\rangle \geq 0$ for all $\psi$ is violated. One way to see this is to compute, using Eq. (1), the initial rate of change of the purity assuming an initially pure state with position uncertainty $\Delta q$ [6–8]:

$$\left[\frac{d\,\mathrm{Tr}\,\rho^2}{dt}\right]_{t=0} = \frac{C\hbar}{4m} - \frac{CkT(\Delta q)^2}{\hbar}. \quad (2)$$

Since this rate may be made positive with a judicious choice of the parameters appearing in the right hand side of Eq. (2), and noting that Eq. (1) preserves the norm and Hermiticity of $\rho$, one is forced to conclude that eigenvalues of $\rho$ can become negative [9]. This is physically unacceptable.

From the work of Lindblad [10] (see also [11]) one could have determined *a priori* that the standard quantum Brownian equation is flawed. Consider the master equation $d\rho/dt = [H, \rho]/\hbar i + \mathcal{L}_{\mathrm{ir}}\rho$ where the dissipator, $\mathcal{L}_{\mathrm{ir}}$, is the operator responsible for irreversible evolution. Lindblad has shown that $[H, \,]/\hbar i + \mathcal{L}_{\mathrm{ir}} \in \mathrm{CD}(\mathcal{H})_\sigma$ if, and only if, $\mathcal{L}_{\mathrm{ir}}$ is of the form $-\sum_\alpha \{A_\alpha, \,, A_\alpha\}$ where the $A_\alpha$ are arbitrary operators [12]. Equation (1) is not of this "Lindblad form" and therefore cannot generally preserve positivity.

Since Lindblad's work appeared, there have been a few efforts to derive well-behaved master equations. However, because most of these have been axiomatic or phenomenological approaches [13], where one typically starts with a master equation already in Lindblad form and then proceeds to place some constraints on the fluctuation and dissipation coefficients, it is not possible to determine whether the resultant equation describes Brownian motion or some other behavior. Indeed, most often the resultant equation corresponds to the positivity-preserving quantum optical master equation [2,14,15] where the secular approximation—that is, the assumption that system periods are much shorter than the coarse-graining time $\Delta t$—is invoked. In contrast, the quantum Brownian equation is obtained by assuming natural periods are much longer than $\Delta t$, the free particle, with an "infinite period," being the quintessential example. In this Letter we derive, within the independent oscillator model, a quantum description of Brownian motion that preserves the norm, Hermiticity, and positivity of $\rho(t)$ for arbitrary system potentials [16].

Let $H = p^2/2m + V(q)$ and $H_r = \sum_j \hbar\omega_j(b_j^{\dagger}b_j + 1/2)$ denote the Hamiltonians of the system of interest and reservoir ($b_j$ is the usual lowering operator), and introduce the coupling $V_T \equiv \hbar q \sum_j \kappa_j(b_j + b_j^{\dagger})$. The independent oscillator model is described by the total Hamiltonian

$$H_T = H + H_r + V_T + q^2\hbar\sum_j \frac{\kappa_j^2}{\omega_j}$$

$$= H + \sum_j \left[\frac{p_j^2}{2m_j} + \frac{k_j}{2}(q_j - q)^2\right], \quad (3)$$

where the last equality holds if $k_j = 2\hbar\kappa_j^2/\omega_j$, $m_j = 2\hbar\kappa_j^2/\omega_j^3$, and $q_j$ and $p_j$ are related in the usual way to $m_j$, $\omega_j$, $b_j$, and $b_j^{\dagger}$. The independent oscillator model has been extensively studied and possesses many attractive features (see, e.g., [17], and references therein). In fact, in Refs. [1,2] this model was used to derive the standard

///

quantum Brownian equation starting from a total initial state that is the product of system of interest and reservoir density operators:

$$\rho(t_0) \otimes \rho_{r,\text{can}}, \qquad (4)$$

where, letting $Z_r$ be the canonical partition function of the reservoir, $\rho_{r,\text{can}} \equiv Z_r^{-1} \exp(-H_r/kT)$. In contrast, the key to our approach below is to start with a correlated initial state. That taking such a state might bode well for the derivation of the quantum Brownian equation can be gleaned from other work [18] investigating initial transients or "jolts." These jolts are boundary layers that arise because the $\omega_{\max} \to \infty$ limit, where $\omega_{\max}$ is the high-frequency cutoff of the reservoir, does not converge uniformly in any interval containing the initial time. Physically, in the uncorrelated initial state (4), the reservoir is not in local equilibrium with the system. Consequently, the $\omega_{\max} \to \infty$ limit is accompanied by large initial impulses that manifest themselves as jolts. These jolts are particularly problematic in the derivation of a quantum Brownian equation because one usually resorts to a weak coupling approximation [2] in which, for the computation of some reservoir traces, one assumes a product state at arbitrary times; however, an initial product state very rapidly becomes entangled as the total system seeks local equilibrium. Thus, it makes more sense to consider from the start correlated states that describe this entanglement.

*Derivation of a positive quantum Brownian equation.*—Instead of (4), the total initial state we consider is

$$\rho_T(t_0) = \frac{e^{-H_T/2kT} e^{H/2kT} \sigma(t_0) \otimes 1_r e^{H/2kT} e^{-H_T/2kT}}{\text{Tr}_T \ e^{-H_T/2kT} e^{H/2kT} \sigma(t_0) \otimes 1_r e^{H/2kT} e^{-H_T/2kT}}, \qquad (5)$$

where $\sigma(t_0)$, serving as the "bare" density operator of the system of interest, involves only system operators. In addition, we require that $\sigma(t_0)$ be Hermitian and positive, and that $e^{H/2kT} \sigma(t_0) e^{H/2kT}$ be of trace class. Note that total equilibrium corresponds to $\sigma \propto \exp(-H/kT)$. From a formal point of view, it is of course impossible to "prove" the correctness of an initial state such as (5); any positive, Hermitian, and unit-trace total initial state is acceptable. It is therefore not surprising that other correlated total initial states have been considered in the literature [5,19–26]. Motivation for the form of our particular $\rho_T(t_0)$ comes from considering the high-temperature linear response of the system of interest from full equilibrium with the reservoir [19, pp. 71 and 72]. Also, although the physics is different, states of the form (5) are somewhat analogous to "local thermodynamic equilibrium" distributions used in the Chapman-Enskog solution of the Boltzmann equation [27].

We first consider the exact evolution of $\rho_T(t_0)$. At later times we have $\rho_T(t) = \rho_{T,\text{un}}/D$ where $D = \text{Tr}_T \ \rho_{T,\text{un}}$ and the unnormalized density operator $\rho_{T,\text{un}}(t)$ is given by

$$e^{-H_T/2kT} e^{(H+H_r)/2kT} \widetilde{U_T}(t, t_0) \sigma(t_0)$$
$$\otimes \rho_{r,\text{can}} \widetilde{U_T}^\dagger(t, t_0) e^{(H+H_r)/2kT} e^{-H_T/2kT}; \qquad (6)$$

$U_T$ is the propagator of the total system corresponding to $H_T$ and the "tilde" is defined by

$$\widetilde{O_T} = e^{-(H+H_r)/2kT} O_T e^{(H+H_r)/2kT}. \qquad (7)$$

We wish to compute the reduced density operator $\rho(t) = \text{Tr}_r \ \rho_{T,\text{un}}(t)/D$, where the "$r$" subscript indicates that the trace is over reservoir variables. Because we restrict ourselves to the high temperature regime, we approximate

$$e^{(H+H_r)/2kT} e^{-H_T/2kT} = e^{-(V_T + q^2 \hbar \sum_j \kappa_j^2/\omega_j)/2kT} \left( 1_T + (2\sqrt{2} \, kT)^{-2} [V_T + q^2 \hbar \sum_j \frac{\kappa_j^2}{\omega_j}, H + H_r] + \cdots \right)$$
$$\approx \exp\left[ -\frac{1}{2kT} \left( V_T + q^2 \hbar \sum_j \frac{\kappa_j^2}{\omega_j} \right) \right]. \qquad (8)$$

Denoting $\langle x_1 x_2 \ldots |$ by $\langle x_r |$, with this approximation we obtain

$$\langle x_r | \rho_{T,\text{un}} | x_r \rangle = \exp\left[ -\frac{1}{2kT} \sum_j \left( -q k_j x_j + q^2 \hbar \frac{\kappa_j^2}{\omega_j} \right) \right] s(t, x_r) \exp\left[ -\frac{1}{2kT} \sum_j \left( -q k_j x_j + q^2 \hbar \frac{\kappa_j^2}{\omega_j} \right) \right], \qquad (9)$$

where $s(t, x_r) \equiv \langle x_r | \widetilde{U_T}(t, t_0) \sigma(t_0) \otimes \rho_{r,\text{can}} \widetilde{U_T}^\dagger(t, t_0) | x_r \rangle$. We will first calculate the coarse-grained, high temperature limit of $\int dx_r \langle x_r | \rho_{T,\text{un}} | x_r \rangle$; the result, when divided by $D$, will be our expression for $\rho(t)$.

In our calculations, the weak coupling approximation that we invoke is

$$\widetilde{U_T}(t, t_0) \sigma(t_0) \otimes \rho_{r,\text{can}} \widetilde{U_T}^\dagger(t, t_0) \approx \sigma(t) \otimes \rho_{r,\text{can}}. \qquad (10)$$

In usual approaches starting with product initial conditions, the weak coupling approximation is made when computing $\rho(t)$ from expression (4) [2]. One expects the weak coupling approximation to be better justified when, as we do here, it is applied to the computation of $\sigma(t)$ under correlated conditions; we will see that our approximate expression for $\sigma(t)$ will not suffer any irreversible jolts. We also assume the Brownian particle to be considerably more massive than those with which it weakly interacts. Accordingly, we treat $W_T = V_T + q^2 \hbar \sum_j \kappa_j^2/\omega_j + p^2/2m$ as a perturbation and introduce the interaction picture propagator $U_T^I(t, t_0) = e^{i(V+H_r)(t-t_0)/\hbar} U_T(t, t_0)$. This propagator satisfies

$$d\widetilde{U_T^I}(\bar{t}, t)/d\bar{t} = \widetilde{W}_T^{\ I}(\bar{t}, t)\widetilde{U_T^I}(\bar{t}, t)/\hbar i, \quad (11)$$

where the superscript "$I$," unless otherwise noted, denotes the following operation:

$$O_T^I(\bar{t}, t) \equiv e^{i(\tilde{V}+H_r)(\bar{t}-t)/\hbar} O_T e^{-i(\tilde{V}+H_r)(\bar{t}-t)/\hbar}. \quad (12)$$

A second order Born expansion of (11) yields

$$\widetilde{U}_T^I(\bar{t}, t) = 1_T + \frac{1}{\hbar i} \int_t^{\bar{t}} dt' \widetilde{W}_T^{\ I}(t', t) + \left(\frac{1}{\hbar i}\right)^2 \int_t^{\bar{t}} dt' \int_t^{t'} dt'' \widetilde{W}_T^{\ I}(t', t)\widetilde{W}_T^{\ I}(t'', t). \quad (13)$$

Define $s^I(\bar{t}, t, x_r) \equiv e^{i\tilde{V}(\bar{t}-t)/\hbar} s(\bar{t}, x_r) e^{-i\tilde{V}^\dagger(\bar{t}-t)/\hbar}$, and let $|n_r\rangle \equiv |n_1 n_2 \ldots\rangle$ denote the number basis of the reservoir. Then Eq. (13), together with the expression

$$s^I(\bar{t}, t, x_r) = \sum_{n_r} \langle x_r | n_r \rangle \langle n_r | \widetilde{U}_T^I(\bar{t}, t)\sigma(t) \otimes \rho_{r,\text{can}} \widetilde{U}_T^{I\,\dagger}(\bar{t}, t) | n_r \rangle \langle n_r | x_r \rangle, \quad (14)$$

which was obtained using the weak coupling approximation [see (10)], may then be used to compute the coarse-grained derivative. Consistently keeping terms up to second order in $m^{-1/2}$ and $\kappa_j$, we obtain

$$\frac{s^I(t+\Delta t, t, x_r) - s^I(t, t, x_r)}{\Delta t} = \int_t^{t+\Delta t} dt' \left\{ \frac{\tilde{p}^{2I}(t', t)}{2m\hbar i} \frac{s(t, x_r)}{\Delta t} + \frac{1}{2\Delta t}[\tilde{q} s(t, x_r)\tilde{q}^\dagger - \tilde{q}^2 s(t, x_r)] \int_t^{t+\Delta t} dt'' \right. \\ \left. \times \sum_j \kappa_j^2 (2N+1) \cos \omega_j(t' - t'') \right\} + \text{H.c.}, \quad (15)$$

where

$$2N + 1 = \coth\left(\frac{\hbar \omega_j}{kT}\right) + \frac{x_j^2 m_j \omega_j}{\hbar} \operatorname{sech}^2\left(\frac{\hbar \omega_j}{2kT}\right). \quad (16)$$

[It is the last term in Eq. (3) that, to highest order, is responsible for extending the limit of integration from $t'$ in Eq. (13) to $t + \Delta t$ in Eq. (15) [28].] Equations (8) and (15) may be used to compute $\frac{d}{dt} \int dx_r \langle x_r | \rho_{T,\text{un}} | x_r \rangle$. For this computation, we approximate sums over reservoir modes by integrals using the usual density of modes, $g(\omega)$, given by $\kappa^2(\omega)g(\omega) = \frac{C}{2\pi}\omega\theta(\omega_{\max} - \omega)\theta(\omega)$, where $\theta$ is the unit step function, $\omega_{\max}$ is a high frequency cutoff, and $C$ is a constant. As is usually done, we also consider the following limits, the order of which is not immaterial: $\hbar/kT\Delta t \to 0$ followed by $\omega_{\max}\Delta t \to \infty$. Thus, when confronted with the integral $\int_t^{t+\Delta t} dt' \int_t^{t+\Delta t} dt'' \sin[\omega_{\max}(t' - t'')]/[\pi(t' - t'')\Delta t]$, for example, we replace it by unity. Defining $\eta = \hbar^2 \omega_{\max}/4\pi k^2 T^2$, $\mathcal{L}_{\text{rev}} = [H, \ ]/\hbar i$, and $\mathcal{L}_{\text{ir}} = -(CkT/2\hbar) \times \{q, \ , q\}$, we obtain

$$\frac{d}{dt}\int dx_r \langle x_r | \rho_{T,\text{un}}(t) | x_r \rangle = e^{\eta \mathcal{L}_{\text{ir}}} \frac{d\sigma(t)}{dt}, \quad (17)$$

where

$$\frac{d\sigma(t)}{dt} = (\tilde{\mathcal{L}}_{\text{rev}} + \tilde{\mathcal{L}}_{\text{ir}})\sigma(t); \quad (18)$$

for superoperators, the tilde means

$$\tilde{\mathcal{L}}\sigma \equiv e^{-H/2kT}[\mathcal{L}(e^{H/2kT}\sigma e^{H/2kT})]e^{-H/2kT} \quad (19)$$

and hence $\tilde{\mathcal{L}}_{\text{rev}} = \mathcal{L}_{\text{rev}}$.

To be consistent, we now apply the same approximations to the computation of $D$, the result of which we should denote by $D_{\text{approx}}$; in other words, $D_{\text{approx}}$ is the trace, over system of interest variables, of $e^{\eta \mathcal{L}_{\text{ir}}}\sigma(t)$, where $\sigma(t)$ is the solution of Eq. (18). This approximate result for $D$ depends on time, although in an exact treatment it would not. As an indication that our approximations are reasonable, we note that $D_{\text{approx}}$ is time independent in the recoilless limit where the Brownian particle's mass is formally set to infinity (or, equivalently, the kinetic energy in $H$ is set to zero). Our central result, describing quantum Brownian evolution, is then

$$\rho(t) = \frac{e^{\eta \mathcal{L}_{\text{ir}}}\sigma(t)}{\operatorname{Tr}[e^{\eta \mathcal{L}_{\text{ir}}}\sigma(t)]}, \quad (20)$$

where $\sigma(t) = e^{t(\tilde{\mathcal{L}}_{\text{rev}} + \tilde{\mathcal{L}}_{\text{ir}})}\sigma(0)$ [29].

We note that when $\sigma = \exp(-H/kT)/Z$, where $Z$ is the canonical partition function of the system, $\rho$ is stationary. In addition, provided $\sigma(0)$ is positive and $e^{H/2kT}\sigma(0)e^{H/2kT}$ is of trace class, $\rho(t)$ is positive since $\mathcal{L}_{\text{ir}}$ is of Lindblad form. It is implicit in the work of Lindblad [10] and more recently Pechukas [30] that a master equation need not be of Lindblad form to preserve positivity. Indeed, our result provides a concrete example since $d\rho/dt$ is not of this form [31]. Finally, we point out that in practice one obtains $\rho(t)$ by first solving Eq. (18); for example, for a free particle where $V(q) = 0$, the equation that has to be solved is

$$\frac{d\sigma}{dt} = \frac{1}{\hbar i}\left[\frac{p^2}{2\bar{m}} + \frac{C\hbar}{4m}(qp + pq) + \frac{C\hbar^2}{4im}\right]\sigma$$
$$+ \text{H.c.} - (CkT/2\hbar)\{q - \gamma ip, \sigma, q - \gamma ip\},$$
(21)

where $\gamma = \hbar/2mkT$ and $\bar{m}^{-1} = m^{-1} + C\hbar^2 i/2kTm^2$. One then forms $\rho(t)$ with the help of Eq. (20). Although we have not computed the full solution $\sigma(t)$ for a harmonic oscillator, we have checked that the short time propagator, at least, gives rise to a Gaussian process.

Primary funding of A. T.'s work was provided by an NSERC Postdoctoral Fellowship.